# Phase Noise and Intensity Noise of the Pulse Train Generated from Mode-locked Lasers in the Demodulation Measurement

Kan Wu, *Student Member, IEEE,* Ping Shum *Senior Member, IEEE*

Equation Chapter 1 Section 1

*Abstract*—The phase noise and intensity noise of a pulse train are theoretically analyzed in the demodulation measurement. The effect of pulse asymmetry is discussed for the first time using Fourier series. Experimentally, photodetectors with different bandwidth and incident power levels are compared to achieve minimum pulse distortion.

## I. Introduction

LOW-noise mode-locked lasers have wide application in various fields including frequency metrology[1], optical sampling[2] and high-precision clock distribution[3], etc. Two main kinds of noise, phase noise and relative intensity noise (RIN) are usually adopted to describe the quality of a pulse train generated by the mode-locked lasers. There have been two important methods developed to measure the phase noise and RIN. One is the analyzer method first proposed by D. von der Linde[4]. The key idea of this method is that the phase noise (or timing jitter) and RIN manifest themselves as sidebands of the main frequency comb in the RF spectrum of a pulse train. Later many researchers further developed this method and gave more general interpretation of the relation between the RF spectrum and the noise of the pulse train[5-8]. This method fully utilizes the information at each harmonic frequency in RF spectrum and is a method specially designed for the noise measurement of a pulse train. However, the drawback of the analyzer method is that a large portion of the dynamic range of the spectrum analyzer is occupied by the carrier power and therefore only a very small portion of the dynamic range is left for noise measurement. Therefore the noise floor of the measurement system is relatively high. The other method is the demodulation method which is widely used in signal process and H. Tsuchida first adopted it to measure the noise of a pulse train[9] and later R. P. Scott provided a detailed description of this method[10]. This method is now used by the commercial signal source analyzers (SSA), e.g., Agilent E5202B[11] and Rohde & Schwarz FSUP26[12], to measure the phase noise and RIN of a periodic signal. This method effectively suppresses the carrier power by using phase locked loop (PLL) to generate a local oscillation and downconvert the target signal to the baseband. Due to the suppression of the carrier power, the whole dynamic range of the analyzer can be used to measure the noise power. However, this method is originally designed for sine wave signal but a pulse train contains many harmonic frequency components. A usual method is to measure the noise of the fundamental repetition frequency. Then question arises: Is the measured phase noise and RIN of the fundamental repetition frequency equal to the noise of the pulse train in the demodulation method? Previous works mainly focus on the analyzer method and there is few works discussing demodulation method.

In this article we analyze the relation of the noise between the pulse train and its frequency components when using demodulation method. Different from the previous work, our analysis is done in time domain and the effect of measurement method is also considered. Besides phase noise, RIN and pulse width fluctuation, pulse asymmetry is included in the model for the first time. Moreover, the effect of photodetector on the measurement results is also experimentally investigated. Section II briefly introduces the mathematical background of the demodulation method. Section III gives the theoretical analysis of the noise in a pulse train. Section IV discusses the effect of photodetector. Section V is the conclusion.

## II. Demodulation method for sine wave signal

We first briefly introduce the mathematical background of the demodulation method. Suppose the input signal is given in Eq.(1), where $\omega_0$ is the carrier angular frequency, $\Delta a$ is the relative intensity noise, $\Delta\varphi$ is the phase noise and $\Delta t = \Delta\varphi/\omega_0$ is the timing jitter. To measure the phase noise, we generate a local oscillation with same frequency and $\pi/2$ delay, that is $\cos(\omega_0 t + \pi/2) = -\sin(\omega_0 t)$. This local oscillation signal is mixed with the input signal and the low frequency part after mixing is filtered out with a low pass filter, see Eq.(2). Thus the phase noise is obtained.

This work is supported in part by the Defense Science and Technology Agency (DSTA) of Singapore.
K. Wu and P. Shum are with the Network Technology Research Centre, Nanyang Technological University, Singapore 637553 (e-mail: wuka0002@e.ntu.edu.sg; EPShum@ntu.edu.sg).

$$V_{in} = (1+\Delta a)\cos(\omega_0 t + \Delta\varphi) = (1+\Delta a)\cos[\omega_0(t+\Delta t)] \quad (1)$$

$$\begin{aligned} V_{mix} &= K(1+\Delta a)\cos(\omega_0 t + \Delta\varphi)\cdot(-\sin\omega_0 t) \\ &= K/2\cdot(1+\Delta a)[\sin\Delta\varphi - \sin(2\omega_0 t + \Delta\varphi)] \\ \text{filter} &+ \text{amplifier} \Rightarrow (1+\Delta a)\sin\Delta\varphi \approx \Delta\varphi \end{aligned} \quad (2)$$

To measure the RIN, one method is to downconvert the signal to an intermediate frequency $\omega_{IF}$ and then split the signal into two paths and multiply together to realize self-mixing. In this way, the measured noise is $2\Delta a$, see Eq.(3). An even simpler method is to consider RIN as a weak amplitude modulation (AM) of a certain carrier frequency. Then the RIN can be extracted by a standard AM envelope demodulator which is a diode cascaded by a grounded capacitor. But this method is not suitable to measure the RIN of a pulse train.

$$\begin{aligned} V_{mix} &= K(1+\Delta a)^2 \cos^2(\omega_{IF} t + \Delta\varphi) \\ &= K/2\cdot(1+\Delta a)^2[1+\cos(2\omega_{IF} t + 2\Delta\varphi)] \\ \text{filter} &+ \text{amplifier} \Rightarrow (1+\Delta a)^2 \approx 1+2\Delta a \end{aligned} \quad (3)$$

### III. NOISE OF A PULSE TRAIN

Suppose the expression of a noiseless pulse train is given by Eq.(4) where $f(t/\tau)$ is the envelope of the single pulse, $\tau$ is the pulse width (full width at half maximum, FWHM), $f_e$ and $f_o$ are even and odd components of $f(t/\tau)$ and $T$ is the period. Then we include the noise and yield Eq.(5) where $\Delta a_n$ is the relative intensity noise, $\Delta t_n$ is the timing jitter and $\Delta\tau_n$ is the pulse width fluctuation.

$$V = \sum_{n=-\infty}^{+\infty} f(\frac{t-nT}{\tau}) = \sum_{n=-\infty}^{+\infty} f_e(\frac{t-nT}{\tau}) + \sum_{n=-\infty}^{+\infty} f_o(\frac{t-nT}{\tau}) \quad (4)$$

$$V = \sum_{n=-\infty}^{+\infty} (1+\Delta a_n) f(\frac{t-nT+\Delta t_n}{\tau + \Delta\tau_n}) \quad (5)$$

Eq.(5) can be simplified with Taylor expansion in Eq.(6) where $\Delta t(t) = \Delta t_n$, $nT - T/2 < t \leq nT + T/2$ is the continuous expression of $\Delta t_n$ and the same definition can be applied to $\Delta\tau(t)$ and $\Delta a(t)$, $\dot{f} = df(x)/dx = \tau df(t/\tau)/dt$. All higher order noise terms have been neglected and only first order noise terms are considered. More details of the derivation of Eq.(6) can be referred to von der Linde's[4] or L. P. Chen's work[7].

$$V = \left(\sum_{n=-\infty}^{+\infty} f(\frac{t-nT}{\tau})\right)(1+\Delta a(t)) + \left(\sum_{n=-\infty}^{+\infty} \dot{f}(\frac{t-nT}{\tau})\right)\frac{\Delta t(t)}{\tau} \\ - \left(\sum_{n=-\infty}^{+\infty} \dot{f}(\frac{t-nT}{\tau})\frac{t-nT}{\tau}\right)\frac{\Delta\tau(t)}{\tau} \quad (6)$$

Remember that $f(t/\tau)$ can be expressed by its even and odd components $f_e$ and $f_o$. Eq.(6) can be separated into two parts with respect to $f_e$ and $f_o$ in Eq.(7) and Eq.(8). Also note that $V_e$ and $V_o$ are not even and odd components of $V$.

$$V_e = \left(\sum_{n=-\infty}^{+\infty} f_e(\frac{t-nT}{\tau})\right)(1+\Delta a(t)) + \left(\sum_{n=-\infty}^{+\infty} \dot{f}_e(\frac{t-nT}{\tau})\right)\frac{\Delta t(t)}{\tau} \\ - \left(\sum_{n=-\infty}^{+\infty} \dot{f}_e(\frac{t-nT}{\tau})\frac{t-nT}{\tau}\right)\frac{\Delta\tau(t)}{\tau} \quad (7)$$

$$V_o = \left(\sum_{n=-\infty}^{+\infty} f_o(\frac{t-nT}{\tau})\right)(1+\Delta a(t)) + \left(\sum_{n=-\infty}^{+\infty} \dot{f}_o(\frac{t-nT}{\tau})\right)\frac{\Delta t(t)}{\tau} \\ - \left(\sum_{n=-\infty}^{+\infty} \dot{f}_o(\frac{t-nT}{\tau})\frac{t-nT}{\tau}\right)\frac{\Delta\tau(t)}{\tau} \quad (8)$$

In the demodulation method, we are more interested in the expression of each frequency component rather than the whole RF spectrum. We then express Eq.(7) with its Fourier series in Eq.(9)-(11) and Eq.(12) where $\omega_0 = 2\pi/T$ is the fundamental repetition angular frequency. The definition and calculation of coefficients $a_k$ and $c_k$ are given in the Appendix.

$$\sum_{n=-\infty}^{+\infty} f_e(\frac{t-nT}{\tau}) = \frac{a_0}{2} + \sum_{k=1}^{+\infty} a_k \cos k\omega_0 t \quad (9)$$

$$\sum_{n=-\infty}^{+\infty} \dot{f}_e(\frac{t-nT}{\tau})\frac{1}{\tau} = -\sum_{k=1}^{+\infty} a_k k\omega_0 \sin k\omega_0 t \quad (10)$$

$$\sum_{n=-\infty}^{+\infty} \dot{f}_e(\frac{t-nT}{\tau})\frac{t-nT}{\tau} = -\frac{a_0}{2} + \sum_{k=1}^{+\infty} (c_k - a_k) \cos k\omega_0 t \quad (11)$$

$$V_e = \frac{a_0}{2}(1+\Delta a + \frac{\Delta\tau}{\tau}) \\ + \sum_{k=1}^{+\infty} [a_k(1+\Delta a) - (c_k - a_k)\frac{\Delta\tau}{\tau}]\cos k\omega_0 t \\ - \sum_{k=1}^{+\infty} a_k k\omega_0 \Delta t \sin k\omega_0 t \quad (12)$$

If the pulse envelope $f(t/\tau)$ is symmetric, i.e., $f_o(t/\tau) = V_o(t/\tau) = 0$ and the pulse width fluctuation is not considered, i.e., $\Delta\tau = 0$, the expression of a noisy pulse train is then given by Eq.(13) where $A_k \approx 1 + \Delta a$ is the amplitude and $\Delta\varphi_k \approx \tan^{-1}(k\omega_0 \Delta t) \approx k\omega_0 \Delta t$ is the phase noise at each harmonic frequency. This result is the same with the one from von der Linde's derivation. The power of RIN will not change with the order of harmonics $k$ whereas the power of phase noise (or timing jitter) will increase with $k^2$. When the demodulation method is applied to measure the fundamental frequency, e.g., $k=1$ and local oscillation is $-\sin(\omega_0 t)$, it is not difficult to find that the phase noise and the RIN of the fundamental frequency represent the noise of the pulse train exactly.

$$V_e = \frac{a_0}{2}(1+\Delta a) + \sum_{k=1}^{+\infty} a_k (1+\Delta a)\cos k\omega_0 t \\ - \sum_{k=1}^{+\infty} a_k k\omega_0 \Delta t \sin k\omega_0 t \\ \approx \frac{a_0}{2}(1+\Delta a) + \sum_{k=1}^{+\infty} A_k \cos(k\omega_0 t + \Delta\varphi_k) \quad (13)$$

Now we include the effect of pulse asymmetry and pulse width fluctuation. The ideal pulse shape generated from mode-locked lasers is either hyperbolic secant or Gaussian. However, a few effects will cause pulse asymmetry, e.g., the self-steepening effect when pulses propagate in the fiber and the tailing effect of photodetector when optical pulse is converted to electrical pulse. We also note that, the optical pulse width is usual very narrow, typically <10ps from a mode-locked laser. However, due to the limited electrical bandwidth of photodetector, the output electrical pulse is much wider after optical-electrical conversion. From this view, photodetection may have the major contribution to the pulse asymmetry and pulse width fluctuation under an electrical measurement system. For asymmetric pulse, the odd component $f_o(t/\tau)$ of pulse shape is no longer zero and the noise terms in Eq.(8) become to affect the measurement results. Similarly, we express Eq.(8) in its Fourier series in Eq.(14)-(16) and Eq.(17). The definition and calculation of coefficients $b_k$ and $d_k$ are given in the Appendix.

$$\sum_{n=-\infty}^{+\infty} f_o(\frac{t-nT}{\tau}) = \sum_{k=1}^{+\infty} b_k \sin k\omega_0 t \quad (14)$$

$$\sum_{n=-\infty}^{+\infty} \dot{f}_o(\frac{t-nT}{\tau})\frac{1}{\tau} = \sum_{k=1}^{+\infty} b_k \omega_0 \cos k\omega_0 t \quad (15)$$

$$\sum_{n=-\infty}^{+\infty} \dot{f}_o(\frac{t-nT}{\tau})\frac{t-nT}{\tau} = \sum_{k=1}^{+\infty} (d_k - b_k)\sin k\omega_0 t \quad (16)$$

$$V_o = \sum_{k=1}^{+\infty} b_k \omega_0 \Delta t \cos k\omega_0 t \\ + \sum_{k=1}^{+\infty}[b_k(1+\Delta a) - (d_k - b_k)\frac{\Delta\tau}{\tau}]\sin k\omega_0 t \quad (17)$$

The final expression of a noisy pulse train can be expressed in Eq.(18) where $I_k$ and $Q_k$ are the amplitude of two orthogonal frequencies respectively and are given by Eq.(19) and Eq.(20).

$$V = V_e + V_o \\ = \frac{a_0}{2}(1+\Delta a + \frac{\Delta\tau}{\tau}) + \sum_{k=1}^{+\infty}(I_k \cos k\omega_0 t - Q_k \sin k\omega_0 t) \quad (18) \\ = \frac{a_0}{2}(1+\Delta a + \frac{\Delta\tau}{\tau}) + \sum_{k=1}^{+\infty} A_k \cos(k\omega_0 t + \Delta\varphi_k)$$

$$I_k = a_k(1+\Delta a) - (c_k - a_k)\frac{\Delta\tau}{\tau} + b_k \omega_0 \Delta t \quad (19)$$

$$Q_k = -b_k(1+\Delta a) + (d_k - b_k)\frac{\Delta\tau}{\tau} + a_k k\omega_0 \Delta t \quad (20)$$

The total amplitude $A_k$ and phase $\Delta\varphi_k$ can then be calculated in Eq.(21) and Eq.(22) where the relation $d\tan^{-1}(x)/dx = 1/(1+x^2)$ is used. Remember that for demodulation measurement of phase noise, what we get is $\sin\Delta\varphi_k$ rather than $\Delta\varphi_k$, as shown in Eq.(23).

$$A_k = \sqrt{I_k^2 + Q_k^2} \\ = \sqrt{(a_k^2+b_k^2)(1+2\Delta a) + (a_k^2 - a_k c_k + b_k^2 - b_k d_k)\frac{2\Delta\tau}{\tau}} \\ \approx \sqrt{a_k^2+b_k^2}\left(1+\Delta a + \frac{a_k^2 - a_k c_k + b_k^2 - b_k d_k}{a_k^2+b_k^2}\frac{\Delta\tau}{\tau}\right) \quad (21) \\ \equiv S_a\left(1+\Delta a + C_{\tau a}\frac{\Delta\tau}{\tau}\right)$$

$$\Delta\varphi_k = \tan^{-1} Q_k/I_k \\ = \tan^{-1}\frac{-b_k(1+\Delta a) + (d_k - b_k)\frac{\Delta\tau}{\tau} + a_k k\omega_0 \Delta t}{a_k(1+\Delta a) - (c_k - a_k)\frac{\Delta\tau}{\tau} + b_k \omega_0 \Delta t} \\ \approx \tan^{-1}\left(-\frac{b_k}{a_k} + (1+\frac{b_k^2}{a_k^2})k\omega_0 \Delta t + \frac{a_k d_k - b_k c_k}{a_k^2}\frac{\Delta\tau}{\tau}\right) \quad (22) \\ \approx \tan^{-1}\left(-\frac{b_k}{a_k}\right) + k\omega_0 \Delta t + \frac{a_k d_k - b_k c_k}{a_k^2+b_k^2}\frac{\Delta\tau}{\tau}$$

$$\sin\Delta\varphi_k = -\sin\left(\tan^{-1}\left(\frac{b_k}{a_k}\right)\right) \\ + \cos\left(\tan^{-1}\left(\frac{b_k}{a_k}\right)\right)\left(k\omega_0 \Delta t + \frac{a_k d_k - b_k c_k}{a_k^2+b_k^2}\frac{\Delta\tau}{\tau}\right) \quad (23) \\ \approx -\frac{b_k}{a_k} + \left(1 - \frac{b_k^2}{2a_k^2}\right)\left(k\omega_0 \Delta t + \frac{a_k d_k - b_k c_k}{a_k^2+b_k^2}\frac{\Delta\tau}{\tau}\right) \\ \equiv -\frac{b_k}{a_k} + S_\varphi\left(k\omega_0 \Delta t + C_{\tau\varphi}\frac{\Delta\tau}{\tau}\right)$$

From Eq.(21) and Eq.(23), the following conclusions can be found:

1) The measured power spectrum density of RIN contains three components: a DC term, a RIN term and a noise coupling term from pulse width fluctuation with a factor of $C_{\tau a}$. Note that the calculation of RIN is related to DC power, so the factor $S_a$ has no contribution;

2) The measured power spectrum density of phase noise contains three components: a DC term, a scaled phase noise term with a factor of $S_\varphi$ and a noise coupling term from pulse width fluctuation with a factor of $S_\varphi C_{\tau\varphi}$;

3) In the demodulation measurement of both phase noise and intensity noise, these two kinds of noise will not couple to each other directly, but if the intensity noise is related to the pulse width fluctuation, the coupling from intensity noise to phase noise still exists.

## IV. EFFECT OF PHOTODETECTOR

The values of the scaling factor $S_\varphi$ and two coupling coefficients $C_{\tau a}$ and $C_{\tau\varphi}$ can be determined by measuring the pulse shape in a high-speed sampling oscilloscope and calculating the coefficients $a_k \sim d_k$. The pulse train under test is generated from a mode-locked laser with a repetition rate of 37.17MHz, a power of 700uW and a pulse width of ~500fs, which is very similar to the one we previously reported[13]. Experimentally, it is found that, photodetector will significantly affect the pulse width and pulse symmetry. In order to verify the effect of photodetector, two photodetectors, a 10GHz high-speed photodetector (Newport 818-BB-35) and a 2GHz low-speed photodetector (Thorlab DET01CFC), are used for comparison. Two power levels, 700uW and ~330uW (using an external 3dB attenuator) are measured. The electrical pulse shape is measured by a 50GHz sampling scope (Agilent infiniium 86100A). The pulse shapes with different types of photodetectors and different incident power levels are shown in Fig.1. The even ($f_e(t)$) and odd ($f_o(t)$) parts are also calculated and plotted.

photodetector and 330uW incident power (c) 2GHz photodetector and 700uW incident power (d) 2GHz photodetector and 330uW incident power

It can be found that, the pulse shape with 10GHz photodetector at high incident power has a very long tail which indicates that the photodetector has been saturated. This long tail cause a relatively large pulse asymmetry and thus large odd part $f_o(t)$. The pulse shape with 10GHz photodetector at low incident power has the minimum tailing effect (e.g., long tail and tail oscillation) and minimum pulse asymmetry. The pulse shapes with 2GHz photodetector at two different incident power levels are almost identical. There is a very large downward overshoot at falling edge and obvious tail oscillation can be observed. These all result in very large pulse asymmetry and thus very large odd part $f_o(t)$.

The corresponding $a(k) \sim d(k)$ are calculated and plotted in Fig.2. The scaling factor $S_\varphi$ and two coupling coefficients $C_{\tau a}$ and $C_{\tau\varphi}$ are calculated accordingly, as shown in Fig.3.

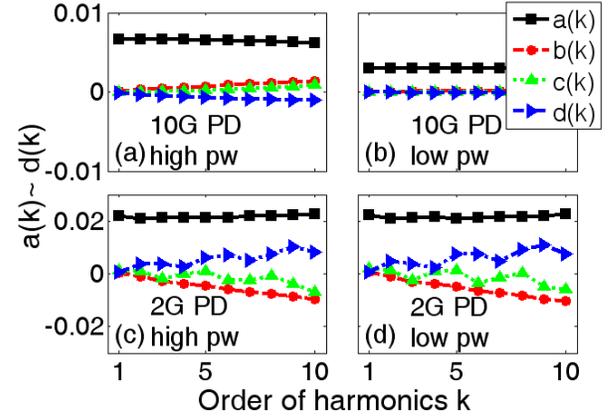

**Fig.2.** The coefficients $a(k) \sim d(k)$ calculated based on the measured pulse shapes in Fig.1. (a) 10GHz photodetector and 700uW incident power (b) 10GHz photodetector and 330uW incident power (c) 2GHz photodetector and 700uW incident power (d) 2GHz photodetector and 330uW incident power

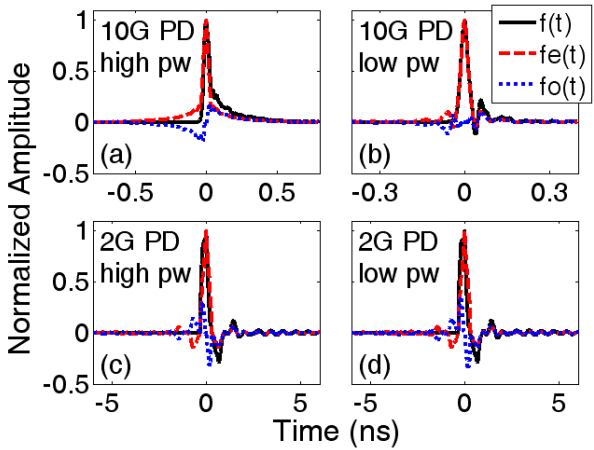

**Fig.1.** Pulse shapes measured by different types of photodetectors at different incident power levels. The even (fe(t)) and odd (fo(t)) parts are also calculated and plotted. (a) 10GHz photodetector and 700uW incident power (b) 10GHz

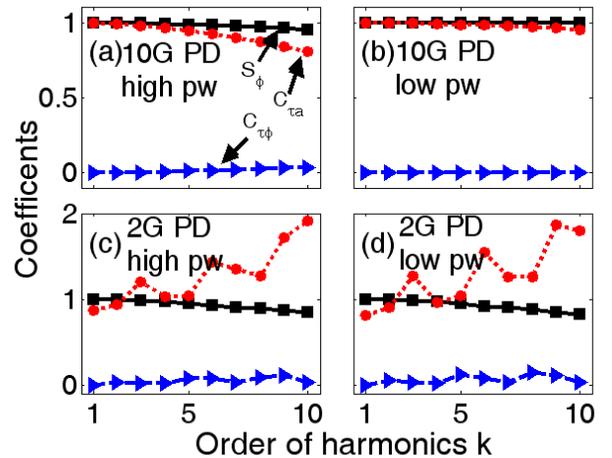

**Fig.3.** The coefficients $S_\varphi$, $C_{\tau a}$ and $C_{\tau\varphi}$ calculated based on the coefficients $a(k) \sim d(k)$ in Fig.2. (a) 10GHz photodetector and 700uW incident power (b) 10GHz photodetector and 330uW incident power (c) 2GHz

photodetector and 700uW incident power (d) 2GHz photodetector and 330uW incident power

Obviously, the pulse measured by 10GHz photodetector has a much better performance with nearly unchanged scaling factor $S_\varphi$ and stable coupling coefficients $C_{\tau a}$ and $C_{\tau \varphi}$ compared with the pulse measured by 2GHz photodetector.

The RF spectra of the pulse trains are also investigated, as shown in Fig.4 and Fig.5. There is a peak at approximate 600MHz~700MHz for 2GHz photodetector cases due to the downward overshoot at falling edge. There are also weak shoulders at ~10kHz offset from the center frequency (not shown in Fig.4 and Fig.5). For 10GHz photodetector, there are additional spurs generated from 0Hz to 1GHz when incident power is high due to the detector saturation. So the pulse train measured by 10GHz photodetector at low incident power has the best measurement distortion with minimum pulse asymmetry, flat RF spectrum and no RF spurs. All these characteristics are summarized in Table 1.

TABLE I
COMPARISON OF PULSE CHARACTERISTICS WITH DIFFERENT PHOTODETECTORS AT DIFFERENT POWER LEVELS

|  | 10G PD High pw | 10G PD Low pw | 2G PD High pw | 2G PD Low pw |
|---|---|---|---|---|
| Pulse width | 45ps | 28ps | 378ps | 380ps |
| Long tail | yes | no | no | no |
| Tail osc. | no | no | yes | yes |
| RF power at 37MHz | -30.1dBm | -42.6dBm | -20.9dBm | -31.6dBm |
| Flat RF spectrum | yes | yes | no | no |
| RF spurs | yes | no | no | no |

Practically, the signal source analyzer (SSA) is usually used to measure the phase noise and RIN with demodulation method. Experimentally, it is found that the equipment has a requirement of input RF power. We measure the phase noise spectra of four cases at the fundamental repetition frequency, as shown in Fig.6 with the SSA (Rohde & Schwarz FSUP26). The shoulders generated by 2GHz photodetectors can be clearly observed and low power case has a higher noise level which suggests that the SSA has a higher system noise with low input power. For 10GHz photodetector cases, two phase noise curves are almost identical except for a difference of noise level which is also due to the SSA's system noise. The RF spurs of the pulse train by 10GHz photodetector at high incident power is not shown in the phase noise spectrum because the spur frequency is far beyond 1MHz region. Since the noise coupling coefficient $C_{\tau\varphi}$ is nearly zero at fundamental frequency for all cases in Fig.3, the results truly represent the phase noise of the pulse trains detected by different photodetectors with different power levels.

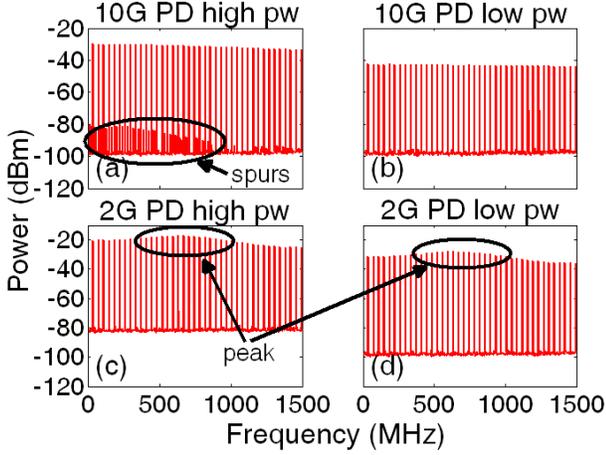

Fig.4. RF spectra of the pulse train measured by different types of photodetectors at different incident power levels with a span from 0Hz to 1.5GHz. (a) 10GHz photodetector and 700uW incident power (b) 10GHz photodetector and 330uW incident power (c) 2GHz photodetector and 700uW incident power (d) 2GHz photodetector and 330uW incident power

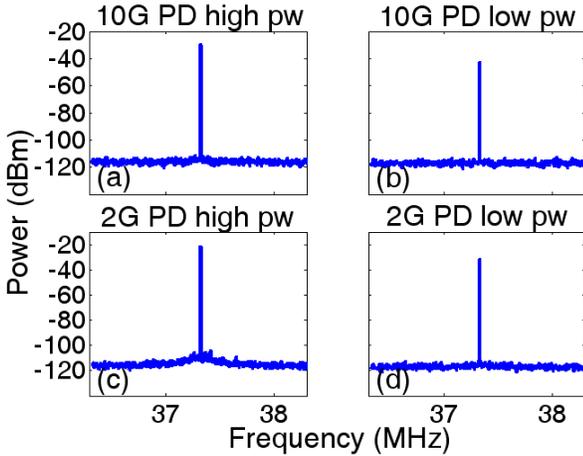

Fig.5. RF spectra of the pulse train measured by different types of photodetectors at different incident power levels with a span of 2MHz centered at the repetition frequency 37.17MHz (a) 10GHz photodetector and 700uW incident power (b) 10GHz photodetector and 330uW incident power (c) 2GHz photodetector and 700uW incident power (d) 2GHz photodetector and 330uW incident power

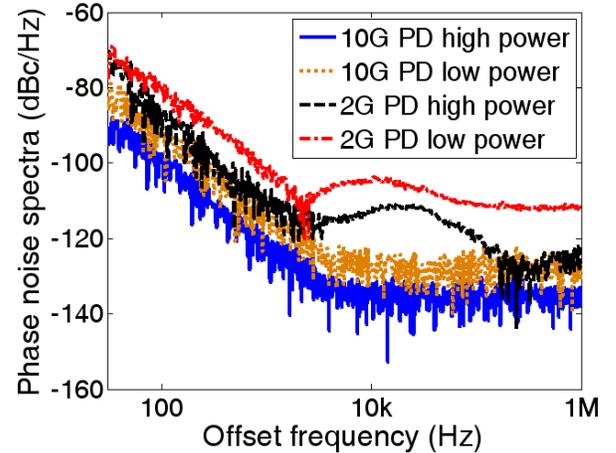

Fig.6. Phase noise spectra of the pulse train measured by different types of photodetectors at different incident power levels

## V.  CONCLUSION

In conclusion, we theoretically investigate the noise of a pulse train using Fourier Series and consider the effect of demodulation measurement. It is the first time that the effect of pulse asymmetry is discussed. It is found that the pulse asymmetry will scale the phase noise by a factor and the pulse width fluctuation will couple to the measurement results of the phase noise and the RIN. Different types of photodetectors with different incident power levels will significantly affect the measurement results. A high speed photodetector with limited incident power will effectively eliminate the pulse distortion e.g., long tail, tail oscillation, RF spurs, etc. and thus generate a reasonable measurement result of the pulse train noise.

## APPENDIX

In the appendix, we give the definition and calculation of the coefficients of Fourier series.

The coefficient $a_k$ of $\sum_{n=-\infty}^{+\infty} f_e(\frac{t-nT}{\tau})$ is defined in Eq.(24) where $x \equiv t/\tau$ and $\sigma \equiv T/\tau$.

$$a_k \equiv \frac{2}{T}\int_{-T/2}^{T/2} f_e(\frac{t}{\tau})\cos k\omega_0 t\, dt \qquad (24)$$
$$= \frac{2}{\sigma}\int_{-\sigma/2}^{\sigma/2} f_e(x)\cos\frac{2\pi kx}{\sigma}dx$$

The coefficient $c_k$ of $\sum_{n=-\infty}^{+\infty} \dot{f}_e(\frac{t-nT}{\tau})\frac{t-nT}{\tau}$ is defined in Eq.(25) and Eq.(26) where we assume the pulse width is much smaller than the period and thus $f_e(\sigma/2)=0$. Also note that $c_0 = 0$.

$$\begin{aligned}
c_k - a_k &\equiv \frac{2}{T}\int_{-T/2}^{T/2} \dot{f}_e(\frac{t}{\tau})\frac{t}{\tau}\cos k\omega_0 t\, dt \\
&= \frac{2}{\sigma}\int_{-\sigma/2}^{\sigma/2} \dot{f}_e(x)x\cos\frac{2\pi kx}{\sigma}dx \\
&= \frac{2}{\sigma}\int_{-\sigma/2}^{\sigma/2} x\cos\frac{2\pi kx}{\sigma}df_e \\
&= \frac{2}{\sigma} f_e(x)x\cos\frac{2\pi kx}{\sigma}\Big|_{-\sigma/2}^{\sigma/2} \\
&\quad -\frac{2}{\sigma}\int_{-\sigma/2}^{\sigma/2} f_e(x)(\cos\frac{2\pi kx}{\sigma} - \frac{2\pi kx}{\sigma}\sin\frac{2\pi kx}{\sigma})dx \\
&= 2f_e(\sigma/2)(-1)^k + \frac{2}{\sigma}\int_{-\sigma/2}^{\sigma/2} f_e(x)\frac{2\pi kx}{\sigma}\sin\frac{2\pi kx}{\sigma}dx \\
&\quad -\frac{2}{\sigma}\int_{-\sigma/2}^{\sigma/2} f_e(x)\cos\frac{2\pi kx}{\sigma}dx
\end{aligned} \qquad (25)$$

$$c_k = \frac{2}{\sigma}\int_{-\sigma/2}^{\sigma/2} f_e(x)\frac{2\pi kx}{\sigma}\sin\frac{2\pi kx}{\sigma}dx \qquad (26)$$

The coefficient $b_k$ of $\sum_{n=-\infty}^{+\infty} f_o(\frac{t-nT}{\tau})$ is defined in Eq.(27).

$$b_k \equiv \frac{2}{T}\int_{-T/2}^{T/2} f_o(\frac{t}{\tau})\sin k\omega_0 t\, dt \qquad (27)$$
$$= \frac{2}{\sigma}\int_{-\sigma/2}^{\sigma/2} f_o(x)\sin\frac{2\pi kx}{\sigma}dx$$

The coefficient $d_k$ of $\sum_{n=-\infty}^{+\infty} \dot{f}_o(\frac{t-nT}{\tau})\frac{t-nT}{\tau}$ is defined in Eq.(28) and Eq.(29)

$$\begin{aligned}
d_k - b_k &\equiv \frac{2}{T}\int_{-T/2}^{T/2} \dot{f}_o(\frac{t}{\tau})\frac{t}{\tau}\sin k\omega_0 t\, dt \\
&= \frac{2}{\sigma}\int_{-\sigma/2}^{\sigma/2} \dot{f}_o(x)x\sin\frac{2\pi kx}{\sigma}dx \\
&= \frac{2}{\sigma}\int_{-\sigma/2}^{\sigma/2} x\sin\frac{2\pi kx}{\sigma}df_o \\
&= \frac{2}{\sigma} f_o(x)x\sin\frac{2\pi kx}{\sigma}\Big|_{-\sigma/2}^{\sigma/2} \\
&\quad -\frac{2}{\sigma}\int_{-\sigma/2}^{\sigma/2} f_o(x)(\sin\frac{2\pi kx}{\sigma} + \frac{2\pi kx}{\sigma}\cos\frac{2\pi kx}{\sigma})dx \\
&= -\frac{2}{\sigma}\int_{-\sigma/2}^{\sigma/2} f_o(x)\frac{2\pi kx}{\sigma}\cos\frac{2\pi kx}{\sigma}dx \\
&\quad -\frac{2}{\sigma}\int_{-\sigma/2}^{\sigma/2} f_o(x)\sin\frac{2\pi kx}{\sigma}dx
\end{aligned} \qquad (28)$$

$$d_k = -\frac{2}{\sigma}\int_{-\sigma/2}^{\sigma/2} f_o(x)\frac{2\pi kx}{\sigma}\cos\frac{2\pi kx}{\sigma}dx \qquad (29)$$

## ACKNOWLEDGMENT

Authors thank the useful discussion from Dr. Chunmei Ouyang, Jia Haur Wong and Dr. Songnian Fu.